\documentclass[conference]{IEEEtran}

\newif\iffull
\fulltrue

\usepackage{algorithmic, algorithm}
\usepackage{amssymb}
\usepackage{amsmath}
\usepackage{xcolor}
\usepackage{cite}
\usepackage{graphicx}
\usepackage{amsthm}
\usepackage{comment}
\usepackage{enumitem}
\setlist[itemize]{leftmargin=*}

\theoremstyle{definition}
\newtheorem{theorem}{Theorem}
\newtheorem{lemma}{Lemma}
\newtheorem{claim}{Claim}
\newtheorem{definition}{Definition}
\newtheorem{corollary}{Corollary}

\newtheorem{construction}{Construction}
\newtheorem{example}{Example}
\newtheorem{remark}{Remark}

\newcommand{\bfe}{{\boldsymbol e}}

\newcommand{\bfw}{{\boldsymbol w}}
\newcommand{\bfx}{{\boldsymbol x}}
\newcommand{\bfy}{{\boldsymbol y}}

\newcommand{\bfalpha}{{\boldsymbol \alpha}}

\newcommand{\cA}{\mathcal{A}}
\newcommand{\cB}{\mathcal{B}}
\newcommand{\cC}{\mathcal{C}}

\newcommand{\cE}{\mathcal{E}}

\newcommand{\cH}{\mathcal{H}}

\newcommand{\cS}{\mathcal{S}}

\renewcommand{\Bbb}{\mathbb}

\newcommand{\N}{{\Bbb N}}

\newcommand{\dg}[1]{{ [{\textcolor{blue!60!black}{#1}} \textcolor{blue!60!black}{--Dganit}]\normalsize}}

\usepackage{tkz-graph}
\usetikzlibrary{arrows}
\usetikzlibrary{positioning}
\usetikzlibrary{calc}
\usepackage{thm-restate}
\usepackage{cleveref}
\usepackage{subcaption}

\usepackage{geometry}
\DeclareRobustCommand*{\IEEEauthorrefmark}[1]{%
  \raisebox{0pt}[0pt][0pt]{\textsuperscript{\footnotesize #1}}%
}

\normalsize

\title{\textbf{{Error-Correcting Codes for Labeled DNA
Sequences}}\vspace{-1ex}}

\author{%
   \IEEEauthorblockN{\textbf{Dganit~Hanania}\IEEEauthorrefmark{1}, 
    and \textbf{Eitan~Yaakobi}\IEEEauthorrefmark{1}}
   \IEEEauthorblockA{\IEEEauthorrefmark{1}%
    Department of Computer Science, 
    Technion---Israel Institute of Technology, Haifa 3200003, Israel}
    
    Email: \{dganit, yaakobi\}@cs.technion.ac.il

\thanks{%
The research was Funded by the European Union (ERC, DNAStorage, 101045114 and EIC, DiDAX 101115134). Views and opinions expressed are however those of the authors only and do not necessarily reflect those of the European Union or the European Research Council Executive Agency. Neither the European Union nor the granting authority can be held responsible for them.}%
\vspace{-2ex}
 }

\IEEEoverridecommandlockouts
\IEEEaftertitletext{\vspace{-5pt}}

\begin{document}

\maketitle

\begin{abstract}
Labeling of DNA molecules is a fundamental technique for DNA visualization and analysis. 
This process was mathematically modeled in~\cite{hanania2023capacity}, where the received sequence indicates the positions of the used labels. 
In this work, we develop error correcting codes for labeled DNA sequences, establishing bounds and constructing explicit systematic encoders for single substitution, insertion, and deletion errors. We focus on two cases: 
(1) using the complete set of length-two labels and (2) using the minimal set of length-two labels that ensures the recovery of DNA sequences from their labeling for ‘almost’
 all DNA sequences.
\end{abstract}

\section{Introduction}
\label{sec:introduction}

Labeling of DNA molecules using fluorescent markers is a key technique in molecular biology and medicine, enabling the visualization and analysis of DNA at the molecular level~\cite{moter2000fluorescence, chen2018efficient, gruszka2021single}. It is widely used in genomics and microbiology for studying gene expression, DNA-protein interactions, and structural variations. 
Techniques such as Fluorescence in situ Hybridization (FISH)~\cite{moter2000fluorescence}, CRISPR~\cite{chen2018efficient,ma2015multicolor}, and Methyltransferases~\cite{deen2017methyltransferase} allow targeting of DNA sequences. 
Labeling can be applied at the per-base level, which is essential for sequencing technologies~\cite{CANARD19941} and epigenomic studies~\cite{mario2002}, or at specific target sequences for applications in species identification\cite{doi:10.3109/1040841X.2016.1169990} and optical mapping~\cite{levy2013beyond, muller2017optical}. This approach also extends to proteins and RNA, supporting molecular analysis~\cite{ohayon2019simulation, alfaro2021emerging} and regulatory studies.

The labeling process was first mathematically modeled in~\cite{hanania2023capacity,hanania2024capacityjournal} as follows. Let $\bfx\in\{A,C,G,T\}^n$ be a DNA sequence and let $\bfalpha\in\{A,C,G,T \}^\ell$ be a label of length $\ell$. The labeling sequence of $\bfx$ indicates the position where the label appears in $\bfx$. 
For example, for $\bfx=ACGTATAGACAC$ and the label $\bfalpha_1=\textcolor{red}{AC}$, the labeling sequence of $\bfx$ is $L_{\bfalpha_1} (\bfx)=\textcolor{red}{1}0000000\textcolor{red}{1}0\textcolor{red}{1}0$. 
This model was extended to accommodate multiple labels. For example, for $\bfalpha_2=\textcolor{blue}{T}$, $L_{\bfalpha_1,\bfalpha_2}(\bfx)=\textcolor{red}{1}00\textcolor{blue}{2}0\textcolor{blue}{2}00\textcolor{red}{1}0\textcolor{red}{1}0$. 
In~\cite{hanania2024capacityjournal}, the authors defined labeling capacity as the maximum information rate achievable using specific patterns as labels on a DNA molecule. 
They also studied the minimum number of labels required to reconstruct the original DNA sequence from its labeling sequence.
For labels of length two, it was shown that ten labels are both sufficient and necessary over the DNA alphabet. 
The authors of~\cite{Hofmeister} extended these results to labels of lengths greater than two. However, the analysis in those papers assumed a perfect labeling and reading process, free of errors.

In contrast to the ideal assumptions made in previous studies, real-world labeling processes are subject to various errors. 
These errors include missed labels (where some labels are undetected due to low efficiency), off-target labeling (where labels bind to unintended sequences), deletion errors (where certain symbols are unreadable by the sequencer), and synchronization errors (where labels are misaligned, shifting left or right in the output).

In this work, we aim to reconstruct the original labeling sequence from a received sequence that may contain errors. We define such error-correcting codes as \emph{error labeling codes} and we focus on cases where the correct labeling sequence uniquely determines the original DNA sequence. 
Our study examines errors in labeled DNA sequences, focusing on two key cases: (1) using a minimal number of length-two labels to achieve maximal capacity and (2) utilizing the complete set of length-two labels.

We specifically address substitution, deletion, and insertion errors, providing both theoretical bounds and explicit constructions for error-correcting codes. 
Our results extend beyond DNA sequences and are applicable to general alphabets.

When considering the full set of labels, we draw connections to the well-studied symbol-pair read channel model~\cite{cassuto2011codes,dinh2017symbol,elishco2019bounds,Khu20} for length-two labels and the $b$-symbol read channel~\cite{yaakobi2016constructions,ding2018maximum,sharma2019b} for longer labels, offering an upper bound and a construction for single deletion or insertion labeling code. For the minimal label set case, we demonstrate the existence of codes that correct a single substitution, as well as those that handle a single deletion or insertion, and present explicit systematic encoding methods.

The rest of this paper is organized as follows. \Cref{sec:defs} formally defines the labeling channel, error labeling codes, and provides key definitions and preliminaries. In \Cref{sec:del}, we present an upper bound for single insertion or deletion labeling codes and provide explicit constructions for both using all labels and a minimal set. In \Cref{sec:sub}, we establish a lower bound on the maximal size of a single substitution labeling code and introduce an explicit encoder for the minimal label set case. Some proofs are omitted due to space limitations.

\section{Definitions and Preliminaries}\label{sec:defs}

Let $\Sigma_q$ denote the $q$-ary alphabet $\{0, 1,\ldots, q-1 \}$. For $q=4$, we mostly refer to the DNA alphabet, that is, ${\Sigma_4= \{A, C, G, T\}}$. 
For a positive integer $n$, let $[n]$ denote the set $\{1, 2, \dots, n\}$.
For a sequence ${\boldsymbol{x}}=(x_1,\ldots,x_n)\in\Sigma_q^n$, and $1\leq i \leq n-k+1$, let ${\boldsymbol{x}}_{[i;k]}= (x_i,\ldots, x_{i+k-1})$. A \emph{label} $\bfalpha\in\Sigma_q^\ell$ is a sequence of (relatively short) length $\ell$ over $\Sigma_q$. We use the \emph{labeling model} as described in~\cite{hanania2024capacityjournal}. 
\vspace{-0.5ex}
\begin{definition}
Let $\bfalpha_1,\ldots,\bfalpha_k$ be $k$ labels of lengths $\ell_1,\ldots,\ell_k$, respectively. Assume no label is a prefix of another label. Denote by $\cA$ the set $\{\bfalpha_1,\ldots,\bfalpha_k\}$, where $\bfalpha_1 \le \dots \le \bfalpha_{\ell}$ are ordered lexicographically.
\begin{itemize}
    \item The $\cA-$\textbf{\emph{labeling sequence}} of  ${\boldsymbol{x}=(x_1,\ldots,x_n)\in\Sigma_q^n}$ is the sequence $L_{\cA}({\boldsymbol{x}}) = (c_1,\ldots,c_n)\in\Sigma_{k+1}^n$, where $c_i=j$ if ${\boldsymbol{x}}_{[i;\ell_j]} = \bfalpha_j$ and $i\leq n-\ell_j+1$, and $c_i=0$ if no such $j$ exists.
    \item A sequence ${\boldsymbol{u}}\in\Sigma^n_{k+1}$ is termed a \textbf{\emph{valid $\cA$-labeling sequence}} if there exists an ${\boldsymbol{x}}\in\Sigma_q^n$ such that ${{\boldsymbol{u}}=L_{\cA}({\boldsymbol{x}})}$.
    \item The \textbf{\emph{labeling capacity}} of $\cA$ is $$\mathsf{cap}(\cA) \triangleq  \limsup_{n\to\infty}\frac{\log_2(|\{L_\cA({\boldsymbol{x}}) : \bfx\in\Sigma_q^n\}|)}{n}.$$
\end{itemize}
\end{definition}

\begin{example}
    For the set of labels $\cA=\{\textcolor{red}{A}, \textcolor{blue}{CC}\}$ and $\bfx= \nolinebreak T\textcolor{red}{A}G\textcolor{blue}{CC}\textcolor{red}{AA}\textcolor{blue}{CCC}G$, it holds that $L_{\cA}=0\textcolor{red}{1}0\textcolor{blue}{2}0\textcolor{red}{11}\textcolor{blue}{22}00$.
\end{example}
 
In~\cite{hanania2024capacityjournal}, the labeling capacity was determined for almost all cases involving a single label, as well as for some cases involving multiple labels. Another problem explored was determining the minimal number of labels of the same length required to achieve the full capacity. 
Specifically, in the context of DNA sequences, the question can be framed as: 
how many labels of a given length $\ell$ are needed to ensure that, given a labeling sequence, the original DNA sequence can (almost always) be reconstructed. 
For \(\ell = 1\), it can be verified that \(q - 1\) labels are required. For \(\ell = 2\), the results for any \(q\) were given in~\cite{hanania2024capacityjournal} by establishing a connection to a graph property where, for any \(k\), there exists at most one path of length \(k\) between any two vertices in a graph with \(q\) vertices.
Such graphs were termed \emph{path-unique graphs}. 
For \(\ell > 2\), the analysis was further developed in~\cite{Hofmeister}, using a similar connection to de Bruijn graphs. The results for \(\ell = 2\) are provided in the following theorem.  

\begin{theorem}(\!~\cite{hanania2024capacityjournal})
    The minimal number of labels of length two over $\Sigma_q$ required to achieve the full labeling capacity is  $\nolinebreak{\phi(q):=q^2-n(q)}$, where
    \vspace{-1ex}
    $$n(q)= \begin{cases}
    \frac{(q+1)^2}{4} & q \text{ is odd}\\
    \frac{q(q+2)}{4} & q \text{ is even}.
\end{cases}$$
\end{theorem}


Specifically, for the DNA case, when $q=4$, ten labels of length two are sufficient and necessary to gain the maximum capacity. 
An example of such set of labels is
\[
\cS = \{AC, CA, GA, GC, GG, GT, TA, TC, TG, TT\}.
\]

In this work, we aim to analyze the use of a set of labels, considering the possibility of errors in the labeling sequence. Our goal is to recover the original labeling sequence from its erroneous version.

\begin{definition}
    Let $\underline{\bfe} = (e_1,e_2,e_3)\in\N^3  $ and let $\cA$ be a set of labels over $\Sigma_q$. For $\bfx\in\Sigma_q^n$, its \emph{$\underline{\bfe}$-error $\cA$-labeling ball}, denoted by $\cB L_{\cA}(\bfx,\underline{\bfe})$, is the set of all possible ${\cA}$-labeling sequences after introducing at most $e_1$ substitution errors, $e_2$ insertions and $e_3$ deletions to $L_{\cA}(\bfx)$. 

    A code $\cC\subseteq\Sigma_q^n$ is called an \emph{$\underline{\bfe}$-error ${\cA}$-labeling code} if for all $\bfx_1,\bfx_2\in\cC$, $\bfx_1\neq\bfx_2$, it holds that $$\cB L_{\cA}(\bfx_1,\underline{\bfe})\cap\cB L_{\cA}(\bfx_2,\underline{\bfe})=\emptyset.
    \vspace{-1ex}$$
    In the case where \( e_2 = e_3 = 0 \), the code is referred to as an \( e_1 \)-substitution \(\mathcal{A}\)-labeling code. Similarly, for \( e_1 = e_3 = 0 \), it is called an \( e_2 \)-insertion \(\mathcal{A}\)-labeling code, and for \( e_1 = e_2 = 0 \), it is referred to as an $e_3$-deletion \(\mathcal{A}\)-labeling code.

\end{definition}
\vspace{-1ex}
Our objective in this work is to construct codes of maximal size for a given $\cA$ and $\underline{\bfe}$. Observe that if the labeling sequences belong to an $\underline{\bfe}$-error correcting code, then the original sequences corresponding to these labeling sequences form an $\underline{\bfe}$-error $\cA$-labeling code. This observation is formalized in the following theorem.

\begin{theorem}\label{known_code}
    Let $\cA$ be a set of labels over $\Sigma_q$. Let $\cC' \subseteq \Sigma_{|\cA|+1}^n$ be an $\underline{\bfe}$-error-correcting code. Then, 
    $$\cC = \{\bfx \in \Sigma_q^n : L_{\cA}(\bfx) \in \cC'\}$$  
    is an $\underline{\bfe}$-error $\cA$-labeling code.  
\end{theorem}

While this theorem provides a useful characterization, it is important to note that not every sequence over $\Sigma_{|\cA|+1}$ is a valid labeling sequence. This adds complexity to the task of finding explicit codes. In this paper, we address this challenge by constructing explicit codes and deriving bounds on their cardinality.  

In this paper, we focus on cases where $\phi(q)$ or more labels of length two are used, which allows sequence reconstruction of all sequences when no labeling errors occur. In this case, correcting errors in the labeling sequence enables recovery of the original sequence. In particular, we provide systematic constructions and bounds for single deletion/insertion $\cA$-labeling codes and single substitution $\cA$-labeling codes. 

\vspace{-1ex}
\begin{remark}\label{known_edges}
In~\cite{hanania2024capacityjournal}, it was shown that full capacity can be achieved using $\phi(q)$ labels. However, ambiguity may arise during recovery of the original sequence if the first or last symbols in the labeling sequence are zero. In such cases, the start or end of the original sequence may not be uniquely determined.
To address this issue, we explicitly fix the symbols before the first and after the last position in the sequence, for example, by setting \( x_0 = x_{n+1} = 0 \). While alternative methods could potentially reduce redundancy, we adopt this approach for its simplicity.
\end{remark}

\begin{remark}
In general, using $|\mathcal{A}|$ labels results in labeling sequences over $\Sigma_{|\mathcal{A}| + 1}$. However, when all labels have the same length $\ell$, i.e., $|\mathcal{A}| = q^\ell$, this is effectively equivalent to using $q^\ell - 1$ labels. In this specific case, the labeling sequences are constructed over the alphabet $\Sigma_{q^\ell}$ rather than $\Sigma_{q^\ell + 1}$.
\end{remark} 
\vspace{-1ex}
The scenario of utilizing all labels of length two corresponds to the well-studied symbol-pair read channel model~\cite{cassuto2011codes,dinh2017symbol,elishco2019bounds,Khu20}, which can be described as follows.
\vspace{-1ex}
\begin{definition}
    Let $\bfx = (x_1, x_2, \dots, x_n) \in \Sigma_q^n$ be a sequence of length $n$ over $\Sigma_q$. The \emph{symbol-pair read sequence of $\bfx$} is  
    $$\pi(\bfx) = ((x_1, x_2), (x_2, x_3), \dots, (x_{n-1}, x_n), (x_n, x_1)) \in \Sigma_{q^2}^n.$$ 
\end{definition}
\vspace{-1ex}
This paradigm closely resembles our framework, with the key distinction being the cyclic reading in the symbol-pair model, where the pair \( (x_n, x_1) \) is included. In contrast, our approach does not assume cyclic reading. Instead, we assume that the symbols before the first and after the last position in the sequence are known, allowing us to complete the labeling sequence in a manner consistent with the cyclic case. Specifically, given a sequence \( \bfx' = x_0 \bfx x_{n+1} \), where \( x_0, x_{n+1} \in \Sigma_q \) are fixed and known, its labeling sequence is  
\[
L_\cA(\bfx') = ((x_0, x_1), \dots, (x_{n-1}, x_n), (x_n, x_{n+1})) \in \Sigma_{q^2}^{n+1}.
\]
The final term \( (x_{n+1}, x_0) \) can be uniquely determined and added to the labeling sequence.

A generalization of the symbol-pair model has also been studied; it involves reading $b$ symbols together and is referred to as the 
$b$-symbol read channel~\cite{yaakobi2016constructions,ding2018maximum,sharma2019b}.
Previous work primarily focused on correcting substitution errors in these models. In~\cite{Khu20}, the problem of synchronization errors was studied, and, in particular, correcting a single deletion in the symbol-pair sequence was addressed. 
However, their solution was restricted to the binary case and did not provide a general bound on the size of the corresponding codes. 
In this paper, we establish a general bound on the code size for the single-deletion case. 
Additionally, we establish the existence of such codes by extending results from the binary case in~\cite{dolecek2010repetition}.

To derive this bound and construct the codes, we introduce the notion of the derivative of a sequence, defined as follows.
\vspace{-1ex}
\begin{definition}
    Let $\bfx = (x_1, \dots, x_n) \in \Sigma_q^n$. The \emph{derivative} of $\bfx$, denoted by $d(\bfx)$, is defined as  
    $$d(\bfx) = (x_1, x_2 - x_1, \dots, x_n - x_{n-1})\in\Sigma_q^n.$$
where all operations are modulo $q$.
\end{definition}

\vspace{-1.5ex}
\section{Single Deletion/Insertion Labeling Codes}\label{sec:del}

In this section, we discuss single insertion or deletion $\cA$-labeling codes for two scenarios: (1) the set of all labels of length two, and (2) a minimal set of labels of length two that achieves full capacity.

It is well known~\cite{levenshtein1966binary} that a code \(\cC\) is a \(t\)-deletion error-correcting code if and only if it is a \(t\)-insertion error-correcting code, and if and only if it is a \((t_1, t_2)\)-error-correcting code for deletions and insertions, respectively, for any \(t_1 + t_2 \leq t\). This property can be also extended for labeling codes.
\vspace{-1ex}
\begin{claim}
    Any code \(\cC\) is a \(t\)-deletion $\cA$-labeling code if and only if it is a \(t\)-insertion $\cA$-labeling code, and if and only if it is a $t_1$-deletion and $t_2$-insertion $\cA$-labeling code, for any \(t_1 + t_2 \leq t\).
\end{claim}
Therefore, we will focus on single-deletion labeling codes, noting that these codes are also capable of correcting a single insertion as well.

\vspace{-0.5ex}
\subsection{Using All Labels of Length Two}
In this section, we first present a construction for single-deletion labeling codes. Then, we derive an upper bound on the size of such codes. Specifically, we focus on labeling sequences obtained using all of the labels of length two.
It can be verified that determining an upper bound on the code size for the case of all labels will directly yield an upper bound for the case involving $\phi(q)$ labels. 

As mentioned earlier, the scenario of using all labels of length two corresponds to the symbol-pair read channel model. 
To construct and bound the single deletion case, we first analyze how a single deletion affects the labeling sequence. 
One part of the following observation was proved in~\cite{Khu20} for the binary pair-symbol channel, and the proof for the other part and the non-binary case is similar.
Before presenting the lemma, we define a zero-deletion error-correcting code as a code capable of correcting the deletion of a zero from its sequences. 
\vspace{-1ex}
\begin{lemma}\label{zero_derivative}
Let $\cA$ be the set of all labels of length two, i.e., $\cA=\Sigma_q^2$. Let $x_0,x_{n+1}\in\Sigma_q$, $\cC'\subseteq \Sigma_q^{n+1}$ and let $$\cC=\left\{\bfx\in\Sigma^n_q: d(\bfx')_{[2;n+1]}\in\cC', \bfx'=x_0\bfx x_{n+1}\right\}.$$ It holds that $\cC'$ is a zero-deletion error correcting code if and only if $\cC$ is a single-deletion $\cA$-labeling code.
\end{lemma}
 
Note that if the first or last label in the labeling sequence is deleted, the deletion can be immediately corrected since \( x_0 \) and \( x_{n+1} \) are assumed to be known.  

Next, we present a single-deletion $\mathcal{A}$-labeling code constructed based on~\Cref{zero_derivative}. Note that any zero-deletion error-correcting code can be used to construct such a single-deletion $\mathcal{A}$-labeling code.
The authors of~\cite{jain2017duplication} characterized codes that address our problem but did not provide explicit constructions for such codes. 
The authors of~\cite{dolecek2010repetition} proposed a code capable of correcting a single insertion of a zero in the binary case. 
Since any code that can correct a single insertion of a zero can also correct a single deletion of a zero~\cite{tallini2019deletion}, this code satisfies our requirements. 
Building on their construction, we extend this method to the non-binary case. 

As defined in~\cite{dolecek2010repetition}, let $A_w^n$ be the set of binary sequences of length $m$ with Hamming weight $w$. Moreover, define 
$$S_{w,a}^{m,w+1}=\left\{(s_1,\dots,s_m)| \sum_{i=1}^{m}i\cdot s_i\equiv a \text{ (mod } w+1)\right\}.$$
The following lemma was proved in~\cite{dolecek2010repetition}.
\vspace{-1ex}
\begin{lemma}\label{dolecek_lemma}
    Each subset $S_{w,a}^{m,w+1}$ is a single zero-insertion correcting code. Moreover,
    there exists $a$ such that $$|S_{w,a}^{m,w+1}|\geq\frac{1}{w+1}{m \choose w}.$$
\end{lemma}

We will use these definitions and lemma in order to construct a single deletion $\cA$-labeling code, when $\cA=\Sigma_q^2$.

\vspace{-1ex}
\begin{theorem}\label{lower_bound}
    There exists a single deletion $\cA$-labeling code $\cC$ such that,
    $$|\cC|\geq \frac{q^{n+1}}{(q-1){(n+2)}}.$$
\end{theorem}

\begin{proof}
Define the following set of sequences of length $m$ over $\Sigma_q$,
\vspace{-1ex}
$$T_{w,a}^{m,w+1}= \left\{(t_1,\dots,t_m):(\lceil \frac{t_1}{q-1}\rceil,\dots,\lceil \frac{t_m}{q-1}\rceil)\in S_{w,a}^{m,w+1}\right\}.$$ From~\Cref{dolecek_lemma}, this is a single zero-insertion
correcting code over $\Sigma_q$.
Since $|T_{w,a}^{m,w+1}|=(q-1)^w\cdot|S_{w,a}^{m,w+1}|$, there exists $a$ such that $$|T_{w,a}^{m,w+1}|\geq\frac{(q-1)^w}{w+1}{m \choose w}.$$
Now, similarly to the proof in~\cite{dolecek2010repetition}, since two non-binary codewords of different Hamming weights cannot result in the same sequence after at most one zero is inserted, we can look at the union over different weights of $T_{w,a^*}^{m,w+1}$, when $a^*=\text{argmax}(|T_{w,a}^{m,w+1}|)$,
\vspace{-1ex}
$$|\cC'|=|\bigcup_{w=0}^m T_{w,a^*}^{m,w+1}|\geq \sum_{w=0}^m\frac{(q-1)^w}{w+1}{m \choose w}=\frac{1}{q-1}\cdot \frac{q^{m+1}}{m+1}.$$
The received set is a single zero insertion (or deletion) correcting code over $\Sigma_q$ for sequences of length $m$.
From~\Cref{zero_derivative}, for $m=n+1$ and any $x_0,x_{n+1}\in\Sigma_q$, $$\cC=\left\{\bfx\in\Sigma^n_q: d(\bfx')_{[2;m]}\in\cC', \bfx'=x_0\bfx x_{n+1}\right\}.$$ is a single deletion $\cA$-labeling code. 
Note that for \( \bfx \in \cC \), we have \( \sum_{i=2}^{n+2} d(x_0\bfx x_{n+1})_i = x_{n+1} - x_0 \). Given \( x_0 \), by the pigeonhole principle, there exists \( x_{n+1} \in \Sigma_q \) such that 
\[
|\cC| \geq \frac{1}{q-1} \cdot \frac{q^{n+2}}{n+2} \cdot \frac{1}{q} = \frac{1}{q-1} \cdot \frac{q^{n+1}}{n+2}.
\]
\end{proof}

\vspace{-2ex}
\begin{remark}
    The explicit cardinality of $S_{w,a}^{m,w+1}$ is being computed in~\cite{dolecek2010repetition}. Since $|T_{w,a}^{m,w+1}|=(q-1)^w\cdot|S_{w,a}^{m,w+1}|$, the cardinality of $T_{w,a}^{m,w+1}$ may be computed as well.
\end{remark}

Next, we will give an upper bound on a single deletion $\cA$-labeling code size. In order to do so, we will use the method of the generalized sphere packing bound, as described in~\cite{fazeli2015generalized, kulkarni2013nonasymptotic}. To discuss our results, some background is given.

We construct a hypergraph $\cH(X,\cE)$ for which the vertices represent sequences of length $m-1$ and the hyperedges are the zero-deletion balls around any sequence of length $m$. More formally, $X=\{\bfx_1,\dots,\bfx_{q^{m-1}}\}=\Sigma_{q}^{m-1}$ and $\cE=\{E_1,\dots,E_{q^{m}}\}=\{B_1^d(\bfx): \bfx\in\Sigma_q^{m}\}$, when $B_1^d(\bfx)$ is the zero-deletion ball of $\bfx$. 
A \emph{transversal} $T\subseteq X$ is a set of vertices that intersects all of the hyperedges in $\cH$. Our goal is to find a minimal size of such a transversal. 
Let $A$ be the $q^{m-1}\times q^{m}$ incidence matrix of $\cH$, that is, $A(i,j)=1$ if $x_i\in E_j$. A transversal can be described as a binary vector $\bfw\in\Sigma_2^{q^{m-1}}$ that satisfies $A^T\cdot\bfw\geq 1$. 
A vector that holds this inequality and its components are values in $\mathbb{R}^+$ is called a \emph{fractional transversal}. 
Our goal is finding such a fractional transversal, since the size of any single zero-deletion correcting code $\cC$, is bounded by $$|\cC|\leq\sum_{\bfx\in\Sigma_{q}^{m-1}}w_\bfx.$$

\begin{theorem}\label{upper_zero_deletion}
    Let $\cC\subseteq \Sigma_q^m$ be a zero-deletion error-correcting code. For $q=2$, it holds that
    $|\cC|\leq \frac{2^{m+2}}{m-2}.$
    For $q>2$,  $|\cC|$ is bounded from above by:
    $$\sum_{z=1}^{\lfloor\frac{m}{2}\rfloor}\sum_{i=0}^{m-2z}{i+z-1 \choose z-1}{m-z-i \choose z}\frac{(q-1)^{m-1-z-i}}{z}.$$
\end{theorem}
\begin{proof}
    For any $\bfx\in\Sigma_q^m$, denote by $z(\bfx)$ the number of zero runs in $\bfx$. Note that $|B_1^d(\bfx)|=z(\bfx)$. Consequently, a property similar to monotonicity holds, meaning that for any $\bfy\in B_1^d(\bfx)$, it follows that $z(\bfy)\leq z(\bfx)$. For this reason, choosing $w_\bfx=\frac{1}{z(\bfx)}$ defines a valid fractional transversal~\cite{fazeli2015generalized}.

When $q=2$, there are $\binom{m+1}{2z}$ sequences of length $m$ over $\Sigma_2$ with exactly $z$ runs of zeroes. Thus, \vspace{-1ex}
$$|\cC|\leq \sum_{x\in\Sigma_{2}^{m-1}}w_x=\sum_{x\in\Sigma_{2}^{m-1}}\frac{1}{z(\bfx)}=\sum_{z=1}^{\lfloor\frac{m}{2}\rfloor}{{m} \choose {2z}}\frac{1}{z}.$$

From~\cite{smagloy2023single, gabrys2015correcting}, it holds that 
\begin{align*}
    \sum_{z=1}^{\lfloor\frac{m}{2}\rfloor}{{m} \choose {2z}}\frac{1}{z}\leq2\sum_{z=1}^{m}{{m} \choose {z}}\frac{1}{z}\leq 2\sum_{j=1}^{m}\frac{2^j-1}{j}\leq\frac{2^{m+2}}{m-2}.
\end{align*}
For the non-binary case, the calculation becomes more complex. In this case, there are $$\sum_{i=0}^{n-(2z-1)}{i+z-1 \choose z-1}{n-(2z-1)-i+z \choose z}(q-1)^{n-z-i}$$ sequences of length $n=m-1$ over $\Sigma_q$ with exactly $z$ runs of zeroes. To explain this, observe that there are $z$ zeroes and $z-1$ non-zero symbols that must appear in such sequences. From the remaining $n-(2z-1)$ symbols, we allocate $i$ zeroes into $z$ runs, and the remainder are divided into $z+1$ non-zero segments. Each non-zero symbol can take any of $q-1$ values.
\end{proof}

\vspace{-1ex}
The following corollary can be derived from~\Cref{zero_derivative} and~\Cref{upper_zero_deletion} by setting \( m = n + 1 \).

\begin{corollary}
    Let if $\cC\subseteq\Sigma_q^n$ be a single-deletion $\cA$-labeling code. For $q=2$, it holds that $|\cC|\leq \frac{2^{n+3}}{n-1}.$
    For $q>2$,  $|\cC|$ is bounded from above by:
    \vspace{-1ex}
    $$\sum_{z=1}^{\lfloor\frac{n+1}{2}\rfloor}\sum_{i=0}^{n+1-2z}{i+z-1 \choose z-1}{n+1-z-i \choose z}\frac{(q-1)^{n-z-i}}{z}.$$
\end{corollary}

We computed the difference between the base-$q$ logarithms of the code size given in~\Cref{lower_bound} and the upper bound established in~\Cref{upper_zero_deletion}. This difference, which reflects the redundancy gap between the construction and the bound, stabilizes around 1 as $n$ increases. These observations indicate that the redundancy of the construction in~\Cref{lower_bound} differs from the upper bound by at most a constant additive gap, which appears to converge to 1.

\subsection{Using Minimal set of Labels of Length Two}
In the previous section, we discussed the scenario where all available labels are used. Now, we turn our attention to a more intricate case: using the minimal number of labels required to achieve full capacity. This scenario is more challenging because deletions or insertions within the labeling sequence can manifest as various types of errors in the original sequence. To address this complexity, we aim to design error-correcting codes for the labeling sequences rather than for the original sequences.

Some of the results in this section will be derived by drawing a connection to known systematic single insertion or deletion error correcting codes. The construction of Varshamov-Tenengolts~\cite{varshamov1965codes} can correct a single insertion or deletion in a binary sequence, and is defined as follows.

\begin{definition}
    For $n>0, a\in\Sigma_{n+1}$, let \vspace{-1ex}$$\text{VT}_a(n)\triangleq \left\{\bfx\in\{0,1\}^n : \sum_{i=1}^nix_i\equiv a \text{ (mod } n+1)\right\}.$$
\end{definition}

Tenengolts presented in~\cite{Tenengolts} a non-binary single insertion or deletion correcting code. In order to present this construction, first we need to introduce the definition of a \emph{signature vector}.

\begin{definition}
Let $\bfx\in\Sigma_q^n$ be a sequence over $\Sigma_q$, when $q>2$. The \emph{signature vector} of $\bfx$ is a binary sequence $s(\bfx)\in\Sigma_2^{n-1}$, where $s(\bfx)_i=1$ for $x_{i+1}\geq x_{i}$ and  $s(\bfx)_i=0$ otherwise for $i\in[n-1]$. 
\end{definition}
\vspace{-1ex}
\begin{definition}
    For $q,n>0, a\in\Sigma_n$ and $b\in\Sigma_q$, let
    \begin{align*}
        \text{T}_{a,b}(n;q)\!\triangleq \! \left\{\!\bfx\in\Sigma_q^n\!: s(\bfx)\!\in\!\text{VT}_a(n\!-\!1), 
    \!\sum_{i=1}^n x_i\!\equiv \!b \text{ mod } q\!  \right\}.
    \end{align*}
\end{definition}

Let $q>0$ and let $\cA$  be a set of $\phi(q)$ labels of length two that achieves full capacity. The following theorem establishes the existence of a single deletion \(\cA\)-labeling code of length \( n \) with a size of at least \( \frac{q^n}{(\phi(q)+1)n} \).

\begin{theorem}\label{exist_T}
     There exist $a\in\Sigma_n,b\in\Sigma_{\phi(q)+1}$ such that $$\cC=\{\bfx\in\Sigma_q^n: L_\cA(\bfx)\in T_{a',b'}(n;\phi(q)+1) \}$$
     is a single deletion $\cA$-labeling code with redundancy of at most $\log_2(n)+\log_2(\phi(q)+1)$ bits.
\end{theorem}

\begin{proof}
    Let $X\subseteq\Sigma^n_q$ be the set of sequences over $\Sigma_q$ of length $n$ and let $L_\cA(X)$ be the set of valid labeling sequences. Denote $p=\phi(q)+1$. 
    For any labeling sequence $\bfy\in L_\cA(X)$, there exists $a\in \Sigma_n, b\in \Sigma_{p}$ such that $\bfy\in T_{a,b}(n;p)$. 
    Since there are $pn$ options to choose $a$ and $b$, from the pigeonhole principle, there exists $a'\in \Sigma_n, b'\in \Sigma_{p}$ for which 
    \begin{align*}
    |T_{a',b'}(n;p)|\geq\frac{|L_\cA(X)|}{pn}\geq \frac{q^n}{pn}= \frac{q^n}{q^{\log_q(pn)}}.
    \vspace{-1.5ex}
    \end{align*}
    
    Since \( T_{a',b'}(n;p) \) corrects a single deletion or insertion error, the proof follows from~\Cref{known_code}.
\end{proof}
\vspace{-1ex}
Next, we will employ Tenengolts' construction~\cite{Tenengolts} to design an explicit encoder. His approach involves appending the syndromes (as redundancy) to the data symbols and using two symbols to separate these parts. 
However, in our case, directly appending redundancy to the end of the sequence may not yield a valid labeling sequence. Hence, a more sophisticated construction is necessary.
The construction presented here is demonstrated using DNA sequences with the label set $\cA = \nolinebreak\{AC, CA, GA, GC, GG, GT, TA, TC, TG, TT\}$. However, it can be adapted to other minimal label sets that achieve maximal labeling capacity and to different alphabet sizes $q$.

Let $\bfx \in \{A,C,G,T\}^k$ and let $\bfy=L_\cA(\bfx)\in\Sigma_{11}^k$ be its labeling sequence. According to Tenengolts' construction, let $s(\bfy)$ be the binary signature vector of $\bfy$. The term $a_{11\rightarrow4}$ represents the conversion of the value $a$ from base $11$ to base $4$. For example, the number $7$ in base $11$ is equivalent to $13$ in base $4$, so $7_{11\rightarrow4}=13$.
\vspace{-0.5ex}
\begin{construction}
The systematic encoder $E_1$ is defined as follows: $$E_1:\{A,C,G,T\}^{k}\rightarrow\{A,C,G,T\}^n,
\vspace{-1ex}$$
$$E_1(\bfx)=(\bfx,s,s,\beta(L_\cA(\bfx))_{11\rightarrow4},\gamma(L_\cA(\bfx))_{11\rightarrow4}),$$
\vspace{-1ex}
when
\vspace{-1ex}
$$ s=
    \begin{cases}
        T \text{ if } x_k=A,C,G\\
        G \text{ if }x_k=T
    \end{cases},
$$
\vspace{-1ex}
$$\beta(\bfy)\triangleq \sum_{i=1}^ky_i \text{ (mod } 11),
\gamma(\bfy)\triangleq\sum_{i=1}^{k-1}i\cdot s(\bfy)_i  \text{ (mod $k$)}.
\vspace{-1ex}$$
\end{construction}

\begin{theorem}
    The code $\cC=\{E_1(\bfx): \bfx\in\Sigma_q^k\}$ is a
    single deletion $\cA$-labeling code, with redundancy of $r=\lceil\log_2(k)\rceil+8 < \lceil\log_2(n)\rceil+ 8$ bits. Furthermore, there exist efficient encoding and decoding algorithms that can correct a single deletion or insertion error, both running in linear time.
\end{theorem}
\vspace{-1ex}
\begin{proof}
In order to prove the theorem, we will provide a decoder.
Let $\bfy'\in\Sigma^{n'}_{11}$ be the received sequence. If $n'=n$, then no deletion or insertion has occurred.

If $n'=n-1$, a deletion occurred. In this case, if $y'_{k}\neq5,10$ (i.e., the $k$th symbol is not the label corresponding to $GG$ or $TT$), then the data part of the sequence remains error-free. We return the DNA sequence corresponding to the first $k$ symbols in $\bfy'$ as their labeling sequence. 
Otherwise, the deletion has occurred in the data part. 
As a result, we can recover the redundancy portion of the DNA sequence from the labeling sequence \(\bfy'_{[k;n-k+1]}\). Let this DNA segment be denoted by \((s, s, \beta, \gamma)\). 
We will convert \(\beta\) and \(\gamma\) to base 11. Since there is no error in this portion, this process allows us to determine \(\beta(\bfy)\) and \(\gamma(\bfy)\), where \(\bfy\) is the original labeling sequence without deletion. 
Using these values, we can correct the deletion in \(\bfy\), according to the Tenengolts' algorithm. Afterwards, we can reconstruct the original DNA sequence from the labeling sequence.

If $n'=n+1$, an insertion has occurred. In this case if $y'_{k+1}= 5,10$, then the data part of the sequence is error-free. Otherwise, the insertion has occurred in the data part. Similarly to the case of deletion, we can recover the syndromes and correct the error in the data part.

The overall time complexity follows from the complexity of Tenengolts' algorithm, combined with the fact that all individual operations involved in encoding and decoding have either constant $O(1)$ or linear $O(n)$ complexity.
\end{proof}

\vspace{-1.5ex}
\section{Single Substitution Labeling Codes}\label{sec:sub}
Substitution errors in the symbol-pair read channel model have been extensively studied~\cite{cassuto2011codes,dinh2017symbol,elishco2019bounds}, so we do not revisit the case of using all labels in this section. Rather, we focus on single substitution $\cA$-labeling codes for a minimal set of length-two labels that achieve full capacity. First, we prove the existence of a code that corrects a single substitution error with redundancy of at most $\log_2(\phi(q)n+1)+\log_2(\phi(q)+1)$.
In the DNA case, where $q=4$, we aim to use a code whose codewords have labeling sequences that belong to the Hamming code over GF($\phi(4)+1$) $=$ GF($11$). Since it is not guaranteed that there are enough valid labeling sequences in the Hamming code, we will use one of the Hamming code cosets. 
Note that the following results apply not only to the DNA case but also to any alphabet size $q$.

Let $\cA\subseteq\Sigma^2_q$ be a minimal set of labels required for achieving full capacity, i.e., $|\cA|=\phi(q)$. Denote the Hamming code over $GF(t)$ by $\cH_t$ and the coset that is received by adding to any codeword the word $\bfy\in\Sigma_t^n$ by $\cH_t+\bfy$. The proof of the following theorem is similar to the proof of~\Cref{exist_T}.
\vspace{-1ex}
\begin{theorem}
    Let $p\geq\phi(q)+1$ be the smallest integer such that there exists a field of size $p$.
    There exists $\bfy\in\Sigma_p^n$ such that
    $$\cC=\{\bfx\in\Sigma_q^n: L_\cA(\bfx)\in \cH_p+\bfy\}$$ is a single substitution $\cA$-labeling code with redundancy of at most $\log_2((p-1)n+1)+\log_2(p)$ bits. 
\end{theorem}

Next, we will present an explicit systematic encoder for DNA sequences. Ideally, we aim to select DNA sequence, determine its labeling sequence, add redundancy using a Hamming code, and convert it back into DNA sequence. 
However, the challenge lies in the fact that the received sequence may not necessarily be a valid labeling sequence. Therefore, our construction will be designed with greater care. The construction is presented for the DNA case but can be adapted to construct codes for different alphabets in a similar manner. Let $\bfx \in \{A,C,G,T\}^k$ and let $\bfy=L_\cA(\bfx)\in\Sigma_{11}^k$ be its labeling sequence.

\begin{construction}
    The systematic encoder $E_2$ is defined as follows: 
    \vspace{-1ex}
    $$E_2:\{A,C,G,T\}^k\rightarrow\{A,C,G,T\}^{n},
    \vspace{-0.8ex}$$
$$E_2(\bfx)=(\bfx,par(\bfy),G,red(\bfy)_{11\rightarrow4}),$$
 when $par(\bfy)$ is the label $\alpha_t\in\cA$ such that  $t=\sum_{i=1}^ky_i\ \text{ (mod } 11)$ and $red(\bfy)$ is the redundancy of $\bfy$ according to $p$-ary Hamming code when $p=11$.
\end{construction}
\vspace{-1ex}
\begin{theorem}
    The code $\cC=\{E_2(\bfx): \bfx\in\Sigma_q^{k}\}$ is a single substitution $\cA$-labeling code, with redundancy of $r =\lceil\log(k+1)\rceil+6< \lceil\log(n+1)\rceil+6$. Furthermore, there exist efficient encoding and decoding algorithms, both running in linear time.
\end{theorem}

\bibliographystyle{IEEEtran}
\bibliography{refs}

\end{document}